\documentclass[12pt,a4paper, dvipdf]{article}
\usepackage [T1]{fontenc} 
\usepackage[ansinew]{inputenc}
\usepackage{epsfig}
\usepackage{amsmath,amsfonts,latexsym, amssymb}
\usepackage{hyperref}
\usepackage{epic}
\usepackage [sort&compress]{natbib}
\bibpunct{[}{]}{,}{n}{}{;} % crochet[5,6], virgule, réf. numerique

\newtheorem {assump}{Assumption}
\DeclareMathOperator {\EE}{{\mathbf {E}}}
\DeclareMathOperator {\sgn}{sgn}
\title{NEW LOOPHOLE FOR THE EINSTEIN-PODOLSKY-ROSEN PARADOX}

\author{Michel Feldmann}
\date{}
\begin{document}

\maketitle

\begin{abstract}
  Using the new concept of "stochastic gauge system", we describe a novel loophole to circumvent the Einstein-Podolsky-Rosen (EPR) paradox. We derive a "realistic" (i.e., classical) model, free from any paradox, which exactly emulates the spin EPR experiment. We conclude that Bell's inequalities are violated in classical physics as well, or, conversely that quantum mechanical theory is logically consistent with relativity.
\end{abstract}

%%%%%%%%%%%%%%%%%%%
\section{INTRODUCTION}

 In 1935, A. Einstein, B. Podolsky, and N. Rosen (EPR) \cite{Einstein}
pointed out a paradox concerning the indetermination principle in quantum mechanics. This paradox was rapidly clarified by N. Bohr \cite{Bohr}. However, in 1952, the EPR questions were reformulated in terms of hidden variables and instantaneous effects at a distance by D. Bohm \cite{Bohm}. Finally, in 1964, J. S. Bell \cite {Bell} demonstrated the so-called Bell theorem, according to which any "local theory" should satisfy a certain inequality and then should violate the standard quantum mechanical theory. Following this paper, a number of experiments were performed to check Bell's inequalities (cf. review papers: Home, 1991 \cite{Home} ; D'Espagnat, 1978 \cite{Despagnat1}; Clauser, 1978 \cite{Clauser1}). The results were non-ambiguous: Bell's inequalities are violated and the predictions of quantum mechanics fully satisfied.

The significance of this phenomenon has been extensively debated (Mermin, 1990 \cite{Mermin}; Clauser, 1969 \cite{Clauser2}; 1974 \cite{Clauser3}; D'Espagnat, 1984 \cite{Despagnat2}; Wigner, 1970 \cite{Wigner}) and several loopholes have been proposed to circumvent the paradox (cf. \cite{Home}).

In the present letter, we suggest a new loophole, which takes advantage of a well-known implicit hypothesis underlying the conventional derivations of Bell's theorem, namely, the "stochastic decoupling" of the probability system from the experimental set-up. When this assumption is rejected, we show that it is feasible to use non-standard probability outcomes, defined as elements of a so-called "stochastic gauge system" : This leads to a possible violation of Bell's inequalities.

The falsity of the "stochastic decoupling" assumption was previously suspected by a number of authors. A significant example of those criticisms was given by Lochak, 1975 \cite{Lochak}. This author claimed that the use of the same probability system in relation to different experimental arrangements is actually in contradiction with quantum mechanical concepts. His argument was founded on the interpretation of a particular event (that we will shortly recall in Sec. 4.2). Nevertheless, his objection was further discarded by Shimony, 1976 \cite{Shimony}
who gave an alternative interpretation of the same event in the framework of the standard theory. In addition, Shimony claimed that the derivation of Bell's inequalities by Clauser and Horne \cite{Clauser3} does not really imply the "stochastic decoupling" assumption. However that may be, in the present letter, we will not focus on this discussion. Instead, we will exhibit for the first time an explicit model in which the assumption is deliberately forsaken. For this, we will replace the conventional probability system by a non standard system of "stochastic gauge variables." Surprisingly, we will find that the framework of classical physics is fully sufficient.

As a matter of fact, it is generally assumed that it is impossible to violate Bell's inequalities in classical physics. But we will just provide such a realistic, (i.e., classical) example which exactly emulates the spin EPR correlations. Nevertheless, at this stage, the loophole has not to be taken as a picture of reality. But at least it proves that Bell's inequalities are violated in classical physics as well, or conversely that quantum mechanical theory is logically consistent with relativity.

When a coupling between the probability system and the experimental set-up is assumed, Bell's inequalities turn out to be irrelevant. However, the need for an instantaneous influence at a distance is not at once eliminated. Clearly, this will be achieved if the probability trial is delayed and performed at a single location between the two ends of the system, so long as all communications of information between such an "ignition point" and the two ends of the system are propagated with a finite velocity. We will prove in Sec. 5 that this is entirely feasible in classical physics provided that some consistency conditions are fulfilled. 

%%%%%%%%%%%%%%%%%%%%%%%%%%%%%%%%%%%%%%%%%%%%%%%%%%%%%%%%%%%%%%%%%%%%%

 \section{
 INCOMPATIBILITY BETWEEN \\QUANTUM MECHANICS AND \\ STANDARD PROBABILITY THEORY}

It is well known that quantum mechanical correlation cannot be regarded as a simple application of standard probability theory. In the particular case of particle spins, this can be proved by the fact that any conventional random system should satisfy Bell's inequalities while spin correlations violate these inequalities.

For clarity, we reformulate here the underlying hypothesis and derive a new proof of Bell's inequalities using the simple concept of "Hamming distance" (see any textbook on digital communications, e.g., Viterbi, 1979 \cite{Viterbi} ).

We consider an ensemble of $N$ pairs of random correlated entities 
$\{S_1, S_2\} $ (e.g., particles). In a space region $ {\cal R}_1$, we suppose that an observer ${\cal O}_1$ selects freely an argument $u_1$ (e.g., a direction of analysis), element of a given set $\mathbf {U}$, and measures on $S_1$ a random dichotomic observable $s_1 = ± 1$ (e.g., a spin), element of a dyadic set $S$ = \{-1, +1\}. Similarly, in a space region ${\cal R}_2$, a second observer ${\cal O}_2$ selects independently an argument $u_2 \in \mathbf {U}$ and, measures on $S_2$ a random dichotomic observable $s_2 =  ± 1 \in S$.

We suppose now that both $u_1$ and $u_2$ remain fixed for the $N$ entity pairs $\{ S_1, S_2 \}$. Hence, a sequence of measurements consists of two observable arrays $ \mathbf{S}_1 = \{ s_{1_1}, s_{1_2} ,..., s_{1_N} \}$ and $ \mathbf{S}_2  = \{ s_{2_1} , s_{2_2} ,..., s_{2_N}\}$. The $N$ measurement pairs can be interpreted as $N$ trials of a random process. Thus, each observable $s_{1_i}$ and $s_{2_i} \in \mathbf{U}$ ($i =1, 2 ,..., N)$ should depend on the arguments $u_1 , u_2 \in \mathbf {U}$ and on the particular outcome $\lambda_i$ of a basic outcome set $\Lambda$ (defined together with its sigma-algebra, $\Sigma (\Lambda )$ and its probability distribution $\rho$) (Kolmogorov, 1956)\cite{Kolmogorov}. The Kolmogorov probability system $\{\Lambda , \Sigma (\Lambda ), \rho \}$ may depend upon the particular arguments $u_1$ and $u_2$. We can write :

\begin{equation}
\label{1}
s_{1_i} = F_1(u_1, u_2, \lambda _i),
\end{equation}

\begin{equation}
\label{2}
s_{2_i} = F_2 (u_1, u_2, \lambda_i),
\end{equation}

where $F_1$ and $F_2$ are measurable mappings of $\Lambda$ onto $ \mathbf {S} = \{ -1,+1 \} $ depending upon arguments $u_1$ and $u_2$ as parameters .

As far as both $u_1$ and $u_2$ remain fixed, owing to the Bernouilli weak law of large numbers, for very large $N$, the arithmetic mean $M(u_1, u_2)$ of the product $s_{1_i} s_{2_i}$ is independent of the actual outcome array $\{\lambda_{i_1}, \lambda_{i_2}, ... \lambda_{i_N}\}$ and equal to the mathematical expectation $ \mathbf {E}[s_1 s_2 ]$ of the observable product with respect to the probability system $\{\Lambda , \Sigma (\Lambda ), \rho \}$.

\begin{equation}
\label{3}
M(u_1, u_2) = (1/N) (s_{1_1} s_{2_1} + s_{1_2} s_{2_2} + ...+ s_{1_N} s_{2_N} ) = \mathbf{E}[s_1 s_2 ].
\end{equation}

It will be convenient to substitute for the observables, $x_i = (1+s_{1_i})/2$ and $y_i = (1+s_{2_i})/2, (i =1, 2,...,N)$. Let $X = \{ x_1, x_2,...x_N \}$ and $Y  =  \{ y_1, y_2,...,y_N \}$. It is seen that $x_i$ and $y_i$ are binary digits (0 or 1) while $X$  and $Y$ are binary words of $N$ bits. In signal processing, the Hamming distance $d_H(X, Y)$ between two N-bit words is defined as the number of homologous bits which are different in the two words $X$ and $Y$. In addition, it is readily shown that $d_H$ has the standard properties of a metrics, i.e., that
$$
d_H(X, Y) \ge 0 ; 
d_H(X, Y) = d_H(Y, X) ; 
d_H(X, Y) = 0 \iff X = Y,
$$
\begin{equation}
\label{4}
|d_H(X, Z) - d_H(Y, Z)| \le d_H(X, Y) \le d_H(X, Z) + d_H(Y, Z).
\end{equation}

The triangle inequality Eq.(\ref{4}), refers to a new arbitrary $N$-bit word $Z$. (Actually, the first part of the triangle inequality is a consequence of the second part).

Coming back to the observable arrays $S_1$ and $S_2$, define

\begin{equation}
\label{5}
d(S_1,S_2) = (1/N) d_H(X, Y).
\end{equation}

From Eq.(\ref{3}), it is a simple exercise to show that

\begin{equation}
\label{6}
d(S_1,S_2) = (1/2) (1- M(u_1, u_2)).
\end{equation}

Presently, any observable array $S_1$ or $S_2$ only pertains to an assigned location, within a pair of random correlated entities $\{S_1, S_2\}$ and to a definite pair of arguments $\{u_1, u_2\}$. In order to check the triangle inequality we have to define a non-ambiguous situation involving three observable arrays $S_a$, $S_b$ and $S_c$ defined at one and the same time, irrespective of their actual environment. For this, we have to force any valid outcome array $\{\lambda_{i_1}, \lambda_{i_2}, ... , \lambda_{i_N}\}$ to fit any argument pair, and hence to assume the following assumption :

\begin{assump}[stochastic decoupling]
\label{decoupling} 
 The Kolmogorov probability system $\{\Lambda , \Sigma (\Lambda ), \rho \}$ is independent of any particular choice of arguments $u_1$ and $u_2$.
\end{assump}
When assumption \ref{decoupling} holds true, a valid $N$-element outcome array $\{\lambda_{i_1},..., \lambda_{i_N}\}$ is simply the collection of the $N$ outcomes from $N$ independent trials performed in $\Lambda$ with respect to the probability distribution $\rho$ .

We will further suppose that the end regions ${\cal R}_1$ and ${\cal R}_2$ are taken very far from one another and with no overlap. Thus, we may assume that the choice of one argument in one region does not influence instantaneously at a distance the experimental result in the second region. This characteristics is usually formalised in terms of "locality principle" :

\begin{assump}[locality]
\label{locality} 

The observable $s_{1_i}$ (resp. $s_{2_i}$) described by Eqs(1)-(2) only depends upon $(u_1, \lambda_i)$ (resp. $(u_2, \lambda_i)$ ), where $u_1, u_2 \in \mathbf{U}$ and $\lambda_i \in \Lambda$.
\end{assump}
Furthermore, to account for spin correlation, the correlation between entities $S_1$ and $S_2$, still governed by Eqs. (\ref{1})-(\ref{2}) is generally specified as follows :

\begin{equation}
\label{7}
\forall \lambda_i \in \Lambda, \forall u_1, u_2 \in \mathbf {U} : u_1= u_2 \Rightarrow s_{1_i} = -s_{2_i}.
\end{equation}

However, for symmetry, and without loss in generality, we will prefer the following definition :

\begin{assump}[correlation]
\label{correlation} 

Let the observables $s_{1_i}$ and $s_{2_i}$ be described by Eqs(\ref{1})-(\ref{2}). We have:

\begin{equation}
\label{8}
\forall \lambda_i \in \Lambda, \forall u_1, u_2 \in \mathbf {U} : u_1= u_2 \Rightarrow s_{1_i} = s_{2_i}.
\end{equation}
\end{assump}

When assumptions \ref{locality} and \ref{correlation} hold true, any observable $s_k$ of an entity $S_k$ depends only upon one argument $u_k \in \mathbf {U}$ and one random outcome $\lambda \in \Lambda$ :

\begin{equation}
\label{9}
s_k = F(u_k, \lambda) ,
\end{equation}

irrespective of $k$ = 1 or 2. Hence, Eq.(\ref{9}) replaces Eqs.\ref{1}-(\ref{2}).

With assumptions 1 to 3, we are able to define without ambiguity three observable arrays, $S_a$, $S_b$ and $S_c$ derived from the same valid outcome array $\{ \lambda_{i_1}, \lambda_{i_2}, ... \lambda_{i_N}\}$, and corresponding respectively to three particular arguments, $a$, $b$ and $c$ of $\mathbf {U}$. Then, clearly, for very large $N$, the distance $d(S_a, S_b)$ between observable arrays, Eq. (\ref{5}), is a mathematical metrics independently of the actual outcome array. From Eq. (\ref{6}), the triangle inequality (Eq.(\ref{4}) first part),

$$|d(S_a, S_c) - d(S_a, S_b)| \ge d(S_b, S_c),$$

is immediately translated into :

\begin{equation}
\label{10}
|M(a, b) - M(a, c)| \le 1 - M(b, c).
\end{equation}
Equation (10) is Bell's inequality, slightly modified to account for the change of sign of $M(b,c)$, since Bell makes use of Eq. (\ref{7}) instead of Eq.(\ref{8}). This inequality is an alternative formulation of the triangle inequality with respect to the Hamming distance.

Now, we measure the spins of a correlated particle pair in directions $a$ and $b$ respectively. Let $s_a$ and $- s_b$ be the results. We repeat this experiment with new particles identically prepared. Let $M(a, b$) be the mean value of the product $s_a \cdot s_b$ for a very large number $N$ of experiments. From quantum mechanical theory, we know that

\begin{equation}
\label{11}
M(a, b)= \cos (a-b).
\end{equation}
This correlation function leads to a violation of Bell's inequalities Eq. (\ref{10}), e.g., for $a = 0, b = \pi/4$ and $c = \pi/2$.
%%%%%%%%%%%%%%%%%%%%%%%%%%%%%%%%%%%%%%%%%%%%%%%%%%%%%%%%%%%%%%%%%%%%
\section {DISCUSSION}
\label{discussion}
We conclude from Sec. 2 that the three assumptions 1-3 do not simultaneously hold true in quantum mechanics. Since the correlation assumption is actually an experimental result, a conventional conclusion is that the locality principle is disproved. However, as the derivation of Bell's inequalities also implies assumption 1 (stochastic decoupling), this last condition is also questionable. Furthermore, if assumption 2 seems rather "plausible", assumption 1 is not evident at all. Above all, as we shall see in Sec. 5, the requirement of instantaneous influence at a distance may be completely eliminated when assumption 1 is rejected.

A simple example of violation of Bell's inequalities when assumptions 2 and 3 hold true but not assumption 1 is sketched in Table 1, in the framework of a discrete 6-element outcome set : The outcome set $\Lambda = \{ \lambda_1, \lambda_2,...\lambda_6\}$ is fixed but the probability distribution, defined as

\begin{equation}
\label{12}
\rho_{{u_1}{u_2}} (\lambda_i) = m p_{u_1} (\lambda_i) + (1-m) p_{u_2} (\lambda_i),
\end{equation}

depends upon the pair of selected arguments $\{u1, u2\}$ in accordance with Table 1. (In Eq.(12), the so-called "parametric probabilities" $p_{u_1} (\lambda_i)$ and $p_{u_2} (\lambda_i)$ are given in Table 1; $i$ =1 to 6 and $u_1, u_2= a, b$ or $c$ ; in addition, $m$ = 0 or 1 is an arbitrary but fixed coefficient). For reasons explained in Sec. 4.2. we will call "stochastic gauge system" such a non standard probability model.

We compute easily from Eqs.(3, 12) and Table 1 :

\begin{equation}
\label{13}
M(a, b) = \mathbf {E} [s_a \cdot s_b ] = \sum_{i=1}^6 s_a(\lambda_i)\cdot s_b(\lambda_i) \cdot \rho_{ab} (\lambda_i) = 2/3 ;
\end{equation}

and similarly, $M (a, c) = 0$ ; $M (b, c) = 2/3$, (regardless of the coefficient $m$ = 0 or 1). Bell's inequality, Eq. (10), is violated.
\begin{table}
$$
\begin{array}{|c|c|c|c|c|c|c|} 
\hline
{}&s_a & s_b & s_c & p_a & p_b & p_c\\
\hline
\lambda_1 &+1&+1&+1&3/12&4/12&3/12\\
\hline
\lambda_2&-1&+1&+1&1/12&1/12&2/12\\
\hline
\lambda_3&-1&-1&+1&2/12&1/12&1/12\\
\hline
\lambda_4&-1&-1&-1&3/12&4/12&3/12\\
\hline
\lambda_5&+1&-1&-1&1/12&1/12&2/12\\
\hline
\lambda_6&+1&+1&-1&2/12&1/12&1/12
\\
\hline
\end{array}
$$

\caption {Example of non standard probability system using "stochastic gauge variables."}
\end{table}
Now, suppose that a first observer ${\cal O}_1$ in an end region ${\cal R}_1$ is allowed to select one of the three arguments $a$, $b$ and $c$. In addition, suppose that the outcome of a random trial is one member of the six element set $\Lambda$ and that the probability distribution $\rho$ involved in $\Lambda$ depends upon the selected argument $u_1 = a, b$ or $c$, according to Eq.(12) and Table 1 (we use $m = 1$ in Eq.(12)). Finally, suppose that the observable is the dichotomic variable $s_{\lambda_i} = F(u_1, \lambda_i)$ also given in Table 1. We will eliminate any instantaneous transmission in Sec. 5, but provisionally assume that the outcome $\lambda_i$ of the random trial is instantaneously transmitted from the end region $ {\cal R}_1$ to a second region $ {\cal R}_2$ where a second observer $ {\cal O}_2$ select independently a second argument $u_2 = a, b$ or $c$. This is exactly the conventional interpretation of EPR experiment which leads to a violation of Bell's inequalities. But before we proceed any further, we must make sure of some consistency conditions.

%%%%%%%%%%%%%%%%%%%%%%%%%%%%%%%%%%
\section {CONSISTENCY CONDITIONS}

\subsection {Derivation of Consistency Conditions}

The key condition in the procedure of Sec 3. is the possibility of performing a definite random trial as soon as one argument $a$ is known (and consequently to ignore the second argument), while the dichotomic observable $s_a$ should be consistent with any argument $b$ as second partner in the pair. The consistency conditions can be formalised as follows:

- A so-called "parametric probability distribution" $p_a$ is associated with each argument $a$. Let $\mathbf {E}_a[.]$ stand for the mathematical expectation associated with this probability distribution pa.

- For any pair $(a, b)$ of arguments, we have :

\begin{equation}
\label{14}
\mathbf{E}_a[s_a ] = \mathbf{E}_b[s_a ] = \mathbf {E}[s_a ]
\end{equation}

\begin{equation}
\label{15}
\mathbf{E}_a[s_a \cdot s_b ] = \mathbf{E}_b[s_a \cdot s_b ]
\end{equation}

Equation (14) may be regarded as a weak version of assumption 1. This defines an overall expectation $\mathbf{E}[s]$ independent of any argument. On the contrary, Eq.(15) is consistent with a very violation of assumption 1.

{\itshape Proof}.  Let $\mathbf{A}^{+} = \{\lambda \in \Lambda / F(a, \lambda) = +1\}$ and similarly for $\mathbf{A}^{-}$, $\mathbf{B}^{+}$ and $\mathbf{B}^{-}$. Since the very observables are dichotomic, we must have a definite probability for, e.g., $\rho_{ab} (\mathbf{A}^{+} )$ and $\rho_{ab} (\mathbf{A}^{+} \bigcap \mathbf{B}^{+} )$ irrespective of the particular value of $m$ in Eq.(12). This is translated in terms of expectations into Eqs.(14-15).

By inspection of Table 1, we see that these conditions are fulfilled.

\subsection {Interpretation in Terms of "Stochastic Gauge Variables"}

It is worth noting that the individual outcomes $\lambda$ are not observable, since they have not a definite probability. Consequently, they should not be interpreted as "hidden variables." In addition, provided that the consistency conditions remain invariant, neither the basic set $\Lambda$ nor the parametric probability distributions $p$ are unique. Thus, the outcomes $\lambda$ may be rather regarded as a form of "stochastic gauge variables."

Incidentally, we will recall an argument given by Lochak, 1975 \cite{Lochak} and discussed by Shimony, 1976 \cite{Shimony}. They consider the event $\mathbf{A}^{+} \bigcap \mathbf{B}^{+}$. As we have seen, this event should have a definite probability. According to Lochak, this is a consequence of assumption 1 (stochastic decoupling). But clearly, this assumption is not necessary. So, we will not debate further of the subtle interpretation of this event.

\subsection {Transmission of information}

We will now consider the possibility of transmitting information from region ${\cal R}_1$ to region ${\cal R}_2$ . Accounting for assumptions 2 and 3 (locality and correlation), we may regard the observable arrays as sequences of digital signals in a memoryless symmetric channel. The input will be one argument, $u_1 = a$ or $b$ in region ${\cal R}_1$ . Let $q_a$ and $q_b$ respectively be the relevant {\itshape a priori} probabilities. This input signal is next encoded by use of a random process as described, e.g., in Table 1, generating a digital sequence. The output will be the observable $s_2$ received on port 2 in the region ${\cal R}_2$ , with the second argument $u_2$ as a parameter.

In ${\cal R}_2$ , in order to discriminate between the two possible input signals, $a$ and $b$, we have to evaluate, e.g., the conditional probability $pr \{u_1 = a / s_2 = +1\}$, for one observable $s_2$ with $u_2$ as a fixed parameter. ($pr \{.\}$ stands for {\itshape "probability that \{.\}"}). By virtue of Eq.(14), we have:

$$
pr\{u_1=a | s_2=1\} = \frac{pr\{s_2 = 1 | u_1=a\} \cdot pr\{ u_1=a\} }{ pr\{s_2=1\} }=pr\{u_1=a \} = q_a.
$$

Irrespective of $u_2$, this probability is identical to the {\itshape a priori} probability. No information can be transmitted.

%%%%%%%%%%%%%%%%%%%%%%%%%%%%%
\section {ELIMINATING INSTANTANEOUS \\ INFLUENCE AT A DISTANCE}
\label{influence}
In this section, we will show that the requirement of instantaneous influence at a distance can be completely eliminated even with a spacelike interval between the two end regions ${\cal R}_1$ and ${\cal R}_2$ . For this, we will emulate an idealised EPR experiment (Fig. 1).

We consider a restframe where a source $S$ and two observers located in regions ${\cal R}_1$  and ${\cal R}_2$  respectively are fixed. In addition, between the two end regions, we define an "ignition point" ${\cal P}$. A two way communication link is established between ${\cal R}_1$  and ${\cal P}$, and ${\cal R}_2$  and ${\cal P}$, respectively. Let $\tau_1$ and $\tau_2$ be the non negative time delays between ${\cal R}_1$-${\cal P}$ and ${\cal R}_2$-${\cal P}$ and let $\tau_1 + \tau_2 = T$.

At time $t_1$, an excited atom is generated from the source.

At time $t_2$, a pair of correlated entities (e.g., photons) is emitted towards the two observers.

At time $t_3$ and $t_4$ respectively, each observer in regions ${\cal R}_1$ and ${\cal R}_2$ opens a polarizer in a random orientation, $u_1$ and $u_2$  respectively, selected among three possibilities, $a$, $b$ and $c$. The two photons reach ${\cal R}_1$ and ${\cal R}_2$ respectively at the same time $t_5$ ($t_5 \ge t_3$ and $t_5 \ge t_4$).

Now, at time $t_5$, the interaction between the photon and the polarizer located in region ${\cal R}_1$ makes the local orientation information $u_1 = a$, $b$ or $c$ to be launched with a finite velocity from ${\cal R}_1$ towards the ignition point ${\cal P}$ (and similarly for the second region ${\cal R}_2$).

At point ${\cal P}$, an information $u_k$ ($k =1$ or 2, e.g., $k = 1$ in Fig. 1) will be first available at time $\tau = t_5 + \tau_k$. The second information, if any, will be ignored. At ${\cal P}$, we perform a random trial with respect to the relevant parametric probability distribution, depending upon this first received argument $u_k$. The outcome of the trial is one member $\lambda$ of the basic set $\Lambda$.

This only outcome $\lambda$ is returned, again with a finite velocity, backwards both end regions ${\cal R}_k$ ($k = 1$ and 2). This information $\lambda$ is received at time $t_6$ and $t_7$ respectively in regions ${\cal R}_1$ and ${\cal R}_2$ where the dichotomic observables $s_k = F(u_k, \lambda)$ are finally determined. 

\begin{figure}
%%%%%%%%%%%%%%%%%%%%%%%%%%%%%%%%%%%%%%%%%%%%%%%%%%%%%%%%%%%%%%%%%%%% figure
\setlength{\unitlength}{12mm}
\begin{picture}(10.5,6)
%légende x=0 
\put(0,5.5){${\cal R}_1$}    %R1
\put(0,0.5){${\cal R}_2$}  %R2
\put(0,4.5){${\cal P}$}      %P
\put(5.6,4.1){${\cal P}$}      %P bis
\put(0,3){S}             %S
%%%%%      légende haut y=5.8
\put(0.6,5.8){$t_1$}       %t1
\put(2.5,5.8){$t_2$}         %t2
\put(4.5,5.8){$t_3$}       %t3
\put(5,5.8){$t_5$}         %t5
\put(6,5.8){$\tau$}        %tau
\put(7,5.8){$t_6$}         %t6
%%%%%%%%
\put(10.2,2.5){$t$}        % t
%%%%%%%% légende bas y=0
\put(0.6,0){$t_1$}
\put(2.5,0){$t_2$}
\put(4.5,0){$t_4$}
\put(5,0){$t_5$}
\put(6,0){$\tau$}
\put(10,0){$t_7$}
%%%%%% lignes horizontales
\put(0.5,4.5){\line(1,0){9.4}}%P
\put(0.5,5.5){\line(1,0){9.4}}%R1
\put(0.5,0.5){\line(1,0){9.4}}%R2
\put(0.5,3){\vector(1,0){9.4}}%S
%%%%% lignes verticales
\put(0.5,0.5){\line(0,1){5}} %t1
\put(2.5,0.5){\line(0,1){5}}   %t2
\put(5,0.5){\line(0,1){5}}   %t5
\put(6,0.5){\line(0,1){5}}   %tau
%%% lignes obliques
\put(5,0.5){\vector(1,1){5}}
%%%%%%%% lignes grasses
\linethickness{0.8mm}
\put(0.5,3){\vector(1,0){2}}%S

\put(2.5,3){\vector(1,1){2.5}}
\put(2.5,2.98){\vector(1,1){2.5}}
\put(2.5,3.02){\vector(1,1){2.5}}
\put(5,5.5){\vector(1,-1){5}}
\put(5,5.48){\vector(1,-1){5}}
\put(5,5.52){\vector(1,-1){5}}
\put(2.5,3){\vector(1,-1){2.5}}
\put(2.5,2.98){\vector(1,-1){2.5}}
\put(2.5,3.02){\vector(1,-1){2.5}}
\put(6,4.5){\vector(1,1){1}}
\put(6,4.48){\vector(1,1){1}}
\put(6,4.52){\vector(1,1){1}}
\end{picture}
%%%%%%%%%%%%%%%%%%%%%%%%%%%%%%%%%%%%%%%%%%%%%%%%%%%%%%%%%%%%%%%%%%%%%%%%%%

\caption {Time-space diagram of an idealised EPR experiment.}
\end{figure}
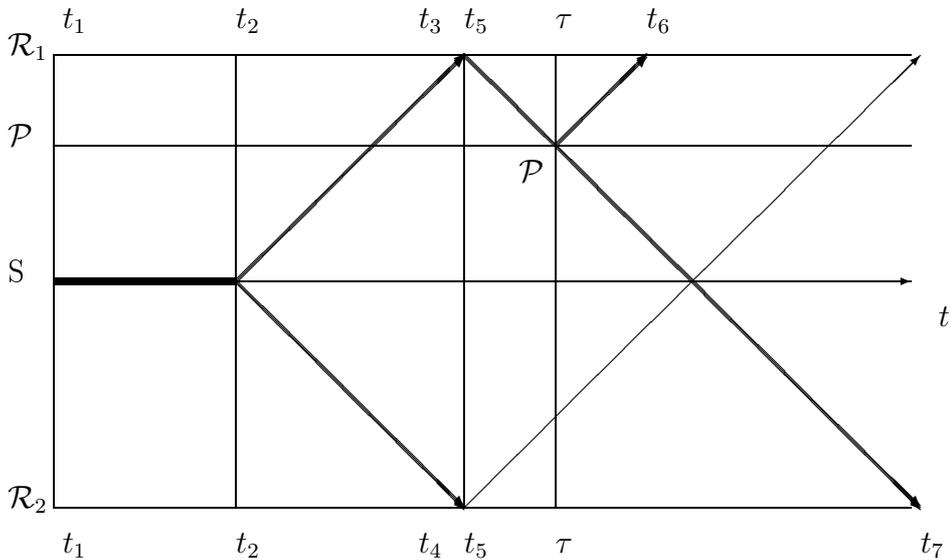
%%%%%%%%%%%%%%%%%%%%%%%%%%%%%%%%%%%%%%%%%%%%%%%%%%%%%%%%%%%%%%%%%%%%%%%%%

Clearly, any communication of information between the end regions and the ignition point will require a finite time lapse and thus will always be consistent with relativity.

The measurements are completed at time $t_6$ and $t_7$. By contrast, the average correlations are immediately defined as soon as the arguments are selected in the end regions, i.e., at time $t_3$ and $t_4$. Therefore, and this statement is surprising but trivial: Statistics are instantaneously definite at a distance. There is no contradiction, because this does not imply any sort of propagation. This just means that if a fact is true, it is instantaneously true everywhere.

In this model, the ignition point ${\cal P}$  will usually be inaccessible and its location should be regarded as an uncontrolled random variable. Therefore, depending upon the location of ${\cal P}$ , $t_6$ and $t_7$ will be randomly distributed within the range $[ t_5, t_5 + T]$. Thus, strictly speaking, for an ensemble of measurements, there is no delay, but rather a finite risetime. The only experimental evidence of such a mechanism will be a risetime dependence, e.g., proportional to the delay $T$ between the two end regions ${\cal R}_1$ and ${\cal R}_2$        .

{\itshape Discussion of the Aspect's experiment.} Most experiments on the EPR paradox only check the violation of Bell's inequalities. According to our model, this does not prove any instantaneous influence at a distance. However, at least one experiment by Aspect, Dalibard and Roger (Aspect, 1982 \cite{Aspect}) seems to have definitely proved the existence of such very instantaneous effects. The authors make use of an acoustooptic switch to redirect the light incident from the source towards one of two polarising cubes in order to forbid any transmission of information with a finite velocity. Actually, our model is much simpler but tentatively, these results may be reinterpreted. An idealised Aspect experiment is very similar to our above description (fig. 1) but in addition, the polarizers are shut at time $t_8 > t_3$, and $t_9 > t_4$ respectively.

Now, according to our model, if $t_3 \le t_5$ and $t_6 \le t_8$ in region ${\cal R}_1$ , and $t_4 \le t_5$ and $t_7 \le t_9$ in region ${\cal R}_2$ , the dichotomic observables will be determined. Otherwise, the photons will not be detected. This will reduce the yield, but this will not affect the statistical distribution. In other words, except from the yield, the experimental results should remain unchanged.

In the actual Aspect's experiment, due to possible leakage of the acoustooptic cells, the times $t_8$ and $t_9$ of total shutting are difficult to appreciate. However, when accounting for the all the delays the yield is found very close to zero. Indeed, the Aspect's yield was unexpectedly low, but no explanation of this fact was reported.

According to our model, it should be emphasised that the crucial effect is not the violation of Bell's inequalities but only the variation of the yield, versus different parameters such as the distance between the two end regions. So far, to the best of our knowledge, no such experiments have been reported.

%%%%%%%%%%%%%%%%%%%%%%%%%%%%%%%%%%%
\section{ A CLASSICAL EMULATION OF EPR EXPERIMENT}
\label{emulation}
The previous example (Sec. \ref{discussion}) may be easily extended to fit the conventional EPR spin experiment. Now, the outcome set will be the unit circle $\Lambda =\{\lambda \}$. We will also allow the arguments $u = a, b$ and $c$ to be any direction of the unit circle $\mathbf {U} =\{u\}$ of the possible spin orientations.

For any argument $u \in {\mathbf U}$ , define a parametric probability density in $\Lambda$ :
$$
p_u (\lambda )d\lambda = (1/4) |\cos(\lambda - u)| d\lambda .$$

Furthermore, define the observable:
$$
F(u, \lambda ) = s_u (\lambda ) = \sgn (\cos(\lambda - u)),$$

where "sgn" stands for $sign$. It is a simple exercise to compute the mathematical parametric expectations :
$$
{\EE}_a[s_a ] =  \int_{0}^{2\pi} s_a (\lambda ) p_a (\lambda ) d\lambda = 0 = {\EE}_b[s_a ] ,
$$

$$
\EE_a[s_a s_b] =  \int_{0}^{2\pi} s_a (\lambda ) s_b (\lambda ) p_a (\lambda ) d\lambda = \cos (a-b) = \EE_b[s_a s_b ] ,
$$

Both consistency conditions Eqs. (14-15) are fulfilled. In addition, the mathematical expectations emulate the conventional spin EPR correlations in accordance with Eq.(11), i.e., $M(a, b) = \cos(a - b)$.

Thus, we can repeat exactly the prior description of a given experiment. The conclusion will be the same : Bell's inequalities are violated. As regards instantaneous influence at a distance, we will assume as in Sec. 5 that the quantum mechanical collapse is equivalent to a stochastic trial occurring at an ignition point ${\cal P}$, located at random between the two end regions. In other words, the two spin directions, or at least one of them, are propagated with a finite velocity from the end regions to the random point ${\cal P}$. This can be interpreted as a very "ignition point" for the quantum mechanical collapse. Next, the collapse front is propagated with a finite velocity from the ignition point backwards the two end regions. Thus, it is possible to forsake instantaneous effects with the only counterpart of a finite risetime depending upon the distance between the two end regions. Nevertheless, the correlations are instantaneously definite at a distance as soon as the spin directions are selected (as explained in Sec. 5).

\section{PHYSICAL INTERPRETATION}

Coming back to the EPR experiment, the quantum state of the system is initially represented by a pure wave function $\Psi_0$. Assume that one argument, (e.g., $u_1 = a$), is selected in one end region, e.g., ${\cal R}_1$. Then, according to our model, the system will be described by the parametric probabilities $p_a$ . As soon as the physical system interacts with region ${\cal R}_1$ the initial state $\Psi_0$ collapses into a mixture of two (equally likely) pure states, $\Psi_1^+$ and $\Psi_1^-$ corresponding respectively to $s_1$ = +1 and $s_1$ = -1.

In region ${\cal R}_1$, let $S_a^1$ be the quantum spin operator of the first particle with respect to the argument $u_1 = a$. The average spin is computed as
$$
\EE_a[s_a]= (1/2) [ <\Psi_1^+ | S_a^1 | \Psi_1^+ > + <\Psi_1^- | S_a^1 | \Psi_1^- > = <\Psi_0 | S_a^1 | \Psi_0 > 
$$
On the other hand, in region ${\cal R}_2$, the second observer selects independently an argument $u_2  = b$. Provisionally, assume that this choice precedes any selection of argument by the first observer in region ${\cal R}_1$. As soon as the physical system interacts with region ${\cal R}_2$,  the initial state $\Psi_0$ collapses into a mixture of two equally likely pure states, $\Psi_2^+$ and $\Psi_2^-$ corresponding respectively to $s_2 = +1$ and $s_2 = -1$. Now, the second observer may evaluate the conditional expectation of the observable $s_1$ if the first argument $u_1$ is assumed to be $u_1 = a$ as

$$
\EE_b[s_a]= (1/2) [ <\Psi_2^+ | S_a^1 | \Psi_2^+ > + <\Psi_2^- | S_a^1 | \Psi_2^- > 
$$

According to quantum mechanical theory, we meet again the first consistency condition,

$$
\EE_a[s_a]=\EE_b[s_a]= <\Psi_0 | S_a^1 | \Psi_0 > 
$$

Now, back in region ${\cal R}_1$ ,we may evaluate the conditional expectation of the product $s_1s_2$  if the second argument $u_2$ is assumed to be $u_2 = b$ as

$$
\EE_a[s_a s_b]= (1/2) [ <\Psi_1^+ | S_a^1  \otimes S_b^2  | \Psi_1^+ > + <\Psi_1^- \otimes  S_b^2 | \Psi_1^- > 
$$

where $S_b^2$ is the quantum spin operator of the second particle with respect to the argument $u_2 = b$.

On the other hand, in region ${\cal R}_2$ , assume that the argument, $u_2 = b$, has been selected. We compute the conditional expectation of the product $s_1 s_2$  if the first argument $u_1$ is assumed to be $u_1 = a$ as
$$
\EE_b[s_a s_b]= (1/2) [ <\Psi_2^+ | S_a^1  \otimes S_b^2  | \Psi_2^+ > + <\Psi_2^- \otimes  S_b^2 | \Psi_2^- > 
$$

From quantum mechanical theory, the second consistency condition holds:

$$
\EE_a[s_a s_b]=\EE_b[s_a s_b]= <\Psi_0 | S_a^1 \otimes S_b^2 | \Psi_0 > 
$$

In conclusion, the parametric probabilities $p_u$ with respect to one argument, $u$, appear to be the probabilities governing the collapse of the initial wave function as soon as one argument $u$ has been selected in one end region.

\section {CONCLUSION}

We have suggested a new loophole to circumvent the EPR paradox. Our model may be sketched as follows : (1) The quantum mechanical collapse is assumed to be propagated with a finite velocity from a random ignition point. (2) Although consistent with quantum mechanical theory, the probability system which governs the collapse depends upon independent arguments to be selected in two spacelike separated regions. (3) As soon as the arguments are selected, the probability system is immediately completed. Therefore, correlations are instantaneously definite at a distance.

We have shown that the only violation of Bell's inequalities does not prove that the quantum mechanical collapse is instantaneous. Our classical model does not aim to describe the underlying reality. However, the experimental signature of a similar quantum mechanical mechanism could be a risetime dependence, e.g., proportional to the distance between the two end regions. Only such experiments, similar to the Aspect's experiment, could check if the collapse is or not definitely instantaneous.

Finally, we have exhibited a non paradoxical realistic model which exactly emulates the spin EPR experiment : This proves that quantum mechanical theory is logically consistent with relativity.

\bibliography{biblio}
\end{document}